\documentclass[aps,showpacs]{revtex4}

\usepackage{amssymb}
\usepackage{graphicx}
\usepackage{bm}
\usepackage[T1]{fontenc}
\usepackage{euscript}
\usepackage{mathrsfs}

\newcommand{\cqg}{Classical Quantum Gravity\ }

\newcommand{\jmp}{J. Math. Phys. (N.Y.)\ }

\begin{document}

\newcommand{\cequ}{\begin{eqnarray}}
\newcommand{\fequ}{\end{eqnarray}}
\newcommand{\anticomut}[2]{\left\{#1,#2\right\}}
\newcommand{\comut}[2]{\left[#1,#2\right]}
\newcommand{\comutd}[2]{\left[#1,#2\right]^{*}}
\newcommand{\density}{{\bf \sigma}}
\newcommand{\pressure}{\mathcal{P}}
\newcommand{\eq}[1]{(\ref{#1})}
\newcommand{\etab}{{\bf \eta}}



\title{Thin-shell wormholes in $d$-dimensional general relativity: \\
Solutions, properties, and stability}

\author{Gon\c{c}alo A. S. Dias\footnote{Email: goncalo.apra@gmail.com}}
\affiliation{Centro de F\'{\i}sica do Porto - CFP\\
Departamento de F\'{\i}sica e Astronomia\\
Faculdade de Ci\^encias da Universidade do Porto - FCUP \\
Rua do Campo Alegre, 4169-007 Porto, Portugal,}
\author{Jos\'e P. S. Lemos\footnote{Email: joselemos@ist.utl.pt}}
\affiliation{Centro Multidisciplinar de {}Astrof\'{\i}sica - CENTRA \\
Departamento de F\'{\i}sica, Instituto Superior T\'ecnico - IST,\\
Universidade T\'ecnica de Lisboa - UTL,\\
Av. Rovisco Pais 1, 1049-001 Lisboa, Portugal \\\&\\
Institute of Theoretical Physics - ITP, Freie Universit\"at Berlin,
Arnimallee 14 D-14195 Berlin, Germany.}


\begin{abstract}

We construct thin-shell electrically charged wormholes in
$d$-dimensional general relativity with a cosmological constant. The
wormholes constructed can have different throat geometries, namely,
spherical, planar and hyperbolic. Unlike the spherical geometry, the
planar and hyperbolic geometries allow for different topologies and in
addition can be interpreted as 
higher-dimensional domain walls or branes connecting two
universes. In the construction we use the cut-and-paste procedure by
joining together two identical vacuum spacetime solutions. Properties
such as the null energy condition and geodesics are studied. A linear
stability analysis around the static solutions is carried out. A
general result for stability is obtained from which previous results
are recovered.
\end{abstract}


\maketitle

\section{Introduction}
\label{introduction}

\subsection{Visser's book}
In physics, in many situations, the study of a system passes through
three stages. First, one finds a solution that emulates the system
itself. Second, in turn, through the solution one studies the
peculiarities that the system might have.  Third, one performs a
stability analysis on the solution to gather whether the system can be
realistic or not.  Traversable wormhole physics is not
different. Throughout its development, traversable wormhole physics
has complied to this bookkeeping.

The field of traversable wormhole physics has been systematized by
Visser in his book ``Lorentzian wormholes: from Einstein to Hawking''
\cite{visserbook}. The three stages of wormhole study are in one form
or another in the book. For instance, not only it displays several
solutions as it also shows how simple wormhole construction can be in
the paper. In addition, the book emphasizes that one of the most
important properties of traversable wormhole physics is that,
within general relativity, its
constitutive matter must possess one kind or another of exotic
properties. Indeed, as the wormhole geometry, with a throat and two
mouths, acts as tunnels from one region of spacetime to another, its
constitutive matter possesses the peculiar property that its
stress-energy tensor violates at least the null energy condition,
i.e., its matter must be exotic. In general, the known classical
fields obey the null energy condition but quantum fields in very
special circumstances may not obey it.  Since the circumstances in
which the null energy condition is not obeyed are very restricted, the
creation of these exotic fields for wormhole building is a difficult
task. In this setting it is thus important to minimize the usage of
exotic fields by finding conditions on the wormhole geometries which
can contain arbitrarily small quantities of matter violating the
averaged null energy condition. This has been clearly explained in
Visser's book \cite{visserbook} and pursued further later
\cite{hocvis}. Finally, stability analysis, perhaps the most difficult
stage of all, is also studied, though timidly, in the book
\cite{visserbook}.

\subsection{Wormhole solutions divided into three classes} The
wormhole solutions that appeared previously have been put into a wider
context through the book \cite{visserbook}, which in turn helped to
inspire developing the field still further as can be seen by the great
many number of solutions dealing with wormhole physics and geometry
that have appeared since. A methodical way of organizing now this
field is by dividing wormhole solutions into three classes:
I. Wormholes generated by continuous fundamental fields with exotic
properties; II. Wormholes generated from matching an interior exotic
solution to an exterior vacuum, at a junction surface, (just like one
does with relativistic stars); and III. Wormholes generated from
exotic thin shells.  Class I could be thought of as nature given, of
cosmological origin. Classes II and III, instead, should be built by
cosmic engineers, first by providing the exotic matter, then building
the wormhole itself and lastly joining it smoothly into the cosmic
vacuum. Class III is the most simple to construct theoretically, and
perhaps also practically.

In relation to class I (wormholes from continuous fundamental fields)
it is striking that the first wormhole solutions are indeed of this
class \cite{hgellisbronnikovtkodamascalarfield}.  Ellis
\cite{hgellisbronnikovtkodamascalarfield} discussed a wormhole
solution, which he called a drainhole, as a model for a particle in
general relativity in which an exotic continuous scalar field through
spacetime provides the means to open up the throat and mouths of the
wormhole. Bronnikov \cite{hgellisbronnikovtkodamascalarfield} found a
general class of solutions within scalar-tensor theories, among which
he managed to glimpse solutions with a neck, as he called it, i.e.,
wormhole solutions.  In the same vein, Kodama
\cite{hgellisbronnikovtkodamascalarfield} studied a wormhole solution,
which he called a kink, with an exotic continuous scalar field being
the necessary field to maintain an Einstein-Rosen bridge open.  Many
other wormholes, whose subsistence hinges on some or another form of
exotic matter put into the model or on the introduction of
interactions of a new type which in turn mimic the exotic matter
itself, have been discussed with profusion.  One can name a few of
those solutions: wormholes in a cosmological constant scenario,
wormhole solutions in semiclassical gravity, wormhole solutions in
Brans-Dicke theory, wormholes on the brane, wormholes supported by
matter with an exotic equation of state such as phantom energy and
tachyon matter, wormholes in nonlinear electrodynamics, wormholes in
nonminimal Einstein--Yang-Mills matter originating a Wu-Yang magnetic
wormhole, wormholes with nonminimal and ghostlike scalar field and
nonminimal matter \cite{wormsmooth}. In this setting, 
cylindrical wormholes, which permit
two definitions of the throat (one of them
allowing non-violation of the energy conditions), with several classes of
fields such as scalar, spinor, Maxwell, and nonlinear electric fields
have also been studied
\cite{wormsmooth}. Stability is always a hard issue and only some of
these systems have had their stability analyzed, see
\cite{bronnikovstabil} as well as some of the works in
\cite{wormsmooth} where for instance the stability of the Ellis
wormhole is studied.

In relation to class II (wormholes generated from matching an interior
exotic solution to an exterior vacuum at a junction surface) there are
many interesting works. Morris and Thorne \cite{mortho} created the
idea that an arbitrarily advanced civilization could construct
wormholes and traverse them to and fro, initiating thus the systematic
study of traversable wormholes, later organized in a coherent whole in
\cite{visserbook}. In Morris and Thorne's work \cite{mortho} a finite
quantity of exotic matter was used in wormhole building. This matter
was then to be carefully matched to the environment. In theory this
matching is done through the junction conditions. In practice, it
certainly can be tricky.  Morris$\,$-Thorne type of wormholes were
later generalized to include a positive or negative cosmological
constant in \cite{llq}, where a review of wormhole solutions was
also taken, and it was suggested that perhaps one should
call a civilization that constructs wormholes an absurdly advanced
civilization rather than an arbitrarily advanced civilization.  Other
wormhole solutions generated from matching an interior exotic solution
to a vacuum, as well as their exotic matter properties, have been
studied. Namely, other wormholes in spacetimes with a cosmological
constant, wormholes on the brane, wormholes made of phantom energy,
wormholes with a generalized Chaplygin gas, Van der Waals quintessence
wormholes \cite{wormjunction}. For some study of stability of
wormholes within this class see \cite{wormjunctionstabil}.

In relation to class III (wormholes generated from exotic thin shells)
many solutions have also been studied. Since thin shells need less
matter, the construction of wormholes from thin shells gives an
elegant way of minimizing the usage of exotic matter. The exotic
matter is concentrated at a shell at the wormhole throat alone,
yielding a thin shell wormhole solution.  In this limiting case of a
thin shell, the throat and the two mouths are all at the
same location, they are indistinguishable.  This concentration of the
entire content of exotic matter at the shell throat of the wormhole is
provided by the cut-and-paste technique used for the first time in
relation to wormholes in \cite{visser12} (see also \cite{visserbook}),
although Morris and Thorne \cite{mortho} had used before heuristic
methods in such a construction.  Thin-shell wormholes in a
cosmological constant background, wormhole with surface stresses on a
thin shell, plane symmetric thin shell wormholes with a negative
cosmological constant, cylindrical thin-shell wormholes, charged
thin-shell Lorentzian wormholes in a dilaton-gravity theory, five
dimensional thin-shell wormholes in Einstein-Maxwell theory with a
Gauss-Bonnet term, solutions in higher dimensional Einstein-Maxwell
theory, thin-shell wormholes associated with global cosmic strings,
wormhole solutions in heterotic string theory, a new type of
thin-shell wormhole by matching two tidal charged black hole solutions
localized on a three brane in the five dimensional Randall-Sundrum
gravity scenario, thin shell and other wormholes within pure
Gauss-Bonnet gravity which can be built without the need of using
exotic matter (the gravity itself being already exotic), spherically
symmetric thin-shell wormholes in a string cloud background spacetime,
and many other solutions have been discussed in detail
\cite{2regions}.  Now, in relation to stability, these thin-shell
wormholes are extremely useful as one may consider a linearized
stability analysis around the static solution.  In the end, the
stability analysis will tell whether the throat can be kept open or
not, i.e., whether the static solution holds under small
perturbations. Within general relativity this has been done for
several thin-shell wormhole systems, namely, four-dimensional
spherical symmetric systems in vacuum \cite{poisson}, with a
cosmological constant \cite{lobolinear}, and with electric charge
\cite{eiroa}, planar systems with a cosmological constant
\cite{lemoslobo}, and $d$-dimensional spherical symmetric systems with
electric charge \cite{rahamanetal}. Also for stability analyses with
dilaton, axion, phantom and other types of matter, and in cylindrical
symmetry see \cite{eeurrrhk}.  This method does not require a specific
equation of state, although it can be added to the analysis as was
done in some works previously cited.  One can also apply other
techniques like a dynamic stability analysis to these thin-shell
wormholes \cite{ishak}.

\subsection{Aim and motivation for this work}

In this work we want to extend further the study of class III
wormholes, wormholes generated from exotic thin shells. We study here
$d$-dimensional thin shell electrically charged wormhole spacetimes
with spherical, planar, and hyperbolic geometry (each with possible
different topologies), with a cosmological constant term within
general relativity.  The motivation for this general analysis comes
from several fronts.

The study of objects in $d$-dimensions was already of interest when
one invoked Kaluza-Klein small extra dimensions within fundamental
theories, and it has become even more interesting and important since
the proposal of the possible existence of a universe with extra large
dimensions \cite{add}. Assuming such a scenario of large extra
dimensions is correct, one can possibly make tiny small distinct
objects such as black holes and wormholes of sizes of the order of the
new Planck scale.

The introduction of a cosmological constant in the study of new
objects is advisable since from astronomical observations, it seems
that we presently live in a world with a positive cosmological
constant, $\Lambda>0$, i.e., in an asymptotically de Sitter universe.
In addition, a spacetime with $\Lambda<0$, an anti-de Sitter
spacetime, is also of significant relevance since it allows a
consistent physical interpretation when one enlarges general
relativity into a gauge extended supergravity theory in which the
vacuum state has negative energy density, i.e., a negative
cosmological constant.  If these theories are correct, they imply that
the anti-de Sitter spacetime should be considered as a symmetric phase
of the theory, although it must have been broken as we do not
presently live in a universe with $\Lambda<0$.  Moreover a $\Lambda<0$
anti-de Sitter spacetime permits a consistent theory of strings in any
dimension, and it has been conjectured that these spacetimes have a
direct correspondence with certain conformal field theories on the
boundary of that space, the AdS-CFT conjecture.  Even with its
preference to negative cosmological constant scenarios, string theory
can, although in a contrived way, produce a landscape of positive
cosmological constant universes, indicating perhaps that one can
transit between both signs of the cosmological constant, see
\cite{lambdaunified} for all these aspects related to the cosmological
constant.

Once one starts to discuss extra dimensions and a cosmological constant
term then the objects under study, black holes or wormholes, say,
can have different geometries
and topologies. Indeed, in the $\Lambda<0$ case,
besides spherical symmetric horizons, black
holes can have planar and hyperbolic symmetric horizons, each one of
these new geometries admitting different topologies \cite{lemosbhs}.
Also, contrary to black holes with a spherical horizon, black holes with
planar and hyperbolic symmetric horizons have the property that
infinity carries the same topology as the throat.  One can go beyond
black hole solutions and build, upon addition of exotic matter,
traversable wormholes with the same corresponding symmetries and
topologies, by natural extension of the corresponding black hole
solutions. In these new planar and hyperbolic geometries, infinity
carries the same topology as the throat, whereas in the spherical
geometry infinity is the usual corresponding asymptotic spacetime. In
this sense, the construction of the planar and hyperbolic symmetric
wormholes does not alter the topology of the background spacetime
(i.e., spacetime is not multiply-connected), so that these solutions
can be considered higher-dimensional domain walls or 
branes.  Note that the $\Lambda=0$ and
$\Lambda<0$ wormhole spacetimes allow valid definitions for the mass
and charge, and when the wormholes have rotation angular momentum is
also well defined \cite{lemosbhs}. So, this gives
further reasons to study, besides wormhole structures in positive
cosmological constant spacetimes, wormhole structures in zero
and negative cosmological constant spacetimes.

Finally, electric charge is a fundamental property of matter, and the
electromagnetic field is a fundamental interaction between
electrically charged objects. Objects with a distribution of electric
charge of the same sign suffer internal repulsion, which counteracts
the effects of attractive gravitation. Moreover, the concept of
wormholes was invented by Wheeler to provide a mechanism for having
charge without charge, since the field lines seen in one part of the
universe could thread the handles of a multiply connected spacetime
and reappear in the other part. This idea was indeed corroborated by
Wheeler and other authors when they showed the Schwarzschild and
Reissner-Nordstr\"om solutions, when fully extended, really should be
interpreted as wormholes, although non-traversable, the latter a
wormhole with electric charge \cite{wheeler}.  As such it is always of
interest to see the effects of a net electric charge in wormholes.

So these are the fields and matter which we consider in our
wormhole and domain wall building. Using the usual cut-and-paste
technique \cite{visserbook} we find, from the vacuum
solutions, thin-shell electrically charged wormholes with different
geometric-topologies in $d$-dimensional general relativity with a
cosmological constant.  We study
the properties of the solutions and analyze perturbations in the
linear regime. Our general solutions encompass previous solutions
which can be reobtained as particular cases of our analysis.
Indeed, previous results in $d=4$ and in other dimensions can be
taken from our general expression, in particular the results in
\cite{poisson,lobolinear,eiroa,lemoslobo,rahamanetal} are reobtained
as particular cases.

\subsection{Structure of the work}
The paper is structured as follows. In
Sec.~\ref{section1} we use the cut-and-paste technique in
order to build static wormholes from the gluing,
above the gravitational radius (or the would-be horizon), of
the  $d$-dimensional electrically charged spacetimes, with a
negative cosmological constant and with several geometries
and topologies. In
Sec.~\ref{ex} we study the properties of these wormholes by providing
the exotic conditions and analyzing the motion of test particles in
the wormhole background. In Sec.~\ref{linearized stability} a
linearized stability analysis is performed around the static
configurations.  Our general result is then shown to recover
previously calculated conditions for static wormholes in particular
dimensions, in particular geometries and topologies, with and without
electric charge, and with and without a cosmological constant. In
Sec. \ref{conclusions} we conclude.  The velocity of light and Newton's
constant are put equal to one throughout.

\section{Spacetime surgery and static wormholes}
\label{section1}
\subsection{Cut-and-paste techniques}
\label{subsection1.1}
\label{subsubsection1.1.1}
We consider Einstein's equations in the form
\begin{equation}
G_{\alpha\beta}+\frac{(d-1)(d-2)}{6}\Lambda\,g_{\alpha\beta} =
8\,\pi\, T_{\alpha\beta}\,,
\label{general einstein}
\end{equation}
where $G_{\alpha\beta}$ is the Einstein tensor, $d$ is the dimension of the
spacetime, $\Lambda$ is the cosmological constant, $g_{\alpha\beta}$ is the
generic metric, and $T_{\alpha\beta}$ is the energy-momentum tensor.
Greek indices are spacetime
indices, latin indices are spacetime indices out of the shell.
Equation (\ref{general einstein}) is supplemented by the
Maxwell equation, but since it is rather trivial, it is not
necessary to explicitly display it.

The general static metric solution for a $d$-dimensional vacuum
spacetime, with electric charge, cosmological constant and different
$(d-2)$ geometric-topologies is given in the form
\begin{equation}
ds^2 = - f(r)dt^2+ f(r)^{-1}dr^2+ r^2d\Omega_{d-2}^k\,,
\label{general metric}
\end{equation}
with the metric function $f(r)$ given by
\begin{equation}
f(r)=k-\frac{\Lambda\, r^2}{3} -
\frac{M}{r^{d-3}}+\frac{Q^2}{r^{2(d-3)}}\,,
\label{metric function}
\end{equation}
and where $\{t,r,\theta_1,\ldots,\theta_{d-2}\}$ are Schwarzschildean
coordinates, $M$ and $Q$ are mass and charge parameters, respectively,
and the cosmological constant $\Lambda$ being zero or of either sign,
$\Lambda<0$, $\Lambda=0$, and $\Lambda>0$.  Here $k$ is the
geometric-topological factor of the $(d-2)$-dimensional $t={\rm
constant}$, $r={\rm constant}$ surfaces, with $k=1,\,0,-1$ for
spherical, planar, and hyperbolic geometries, respectively. The
spherical geometry can only have $(d-2)$-dimensional surfaces with
spherical topology and with infinity behaving normally.  On the other
hand, the $(d-2)$-dimensional hyperbolic and planar geometries can
have different topologies, depending on whether they are infinite in
extent or somehow compactified to produce different
$(d-2)$-dimensional surfaces, and with infinity following the topology
of those $(d-2)$-dimensional surfaces.  For instance, in $d=4$ the
situation simplifies, the $t={\rm constant}$, $r={\rm constant}$
surfaces are 2-surfaces and can be classified.  For spherical symmetry
each 2-surface is compact (genus zero) spherical surface.  In planar
geometry each 2-surface can be an infinite plane, a cylindrical
surface, or a compact (genus one) toroidal surface. In hyperbolic
geometry each 2-surface can be an infinite hyperbolic plane, or a
compact (genus two or greater) toroidal surface with several holes.
For surfaces in higher dimensions the topologies can be even a lot
more complicated \cite{lemosbhs}.  Now, $(d\Omega_{d-2}^k)^2$ is given
by three different expressions according to the value of the parameter
$k$,
\begin{eqnarray}
\label{angular metric differentials}
d\Omega_{d-2}^{1} &=& d\theta_1^2+\sin
\theta_1^2\,d\theta_2^2+\ldots+
\prod_{i=2}^{d-3}\sin\theta_i^2\,d\theta_{d-2}^2\,,\nonumber\\
d\Omega_{d-2}^{0} &=&
d\theta_1^2+d\theta_2^2+\ldots+d\theta_{d-2}^2\,,\nonumber\\
d\Omega_{d-2}^{-1} &=& d\theta_1^2+(\sinh
\theta_1)^2\,d\theta_2^2+\ldots+(\sinh \theta_1)^2
\prod_{i=2}^{d-3}\sin\theta_i^2\,d\theta_{d-2}^2\,.
\end{eqnarray}
One should note that the mass and charge terms above in
Eq.~(\ref{metric function}), that is $M$ and $Q$, are not the ADM mass
and charge of the solutions, but rather $M$ and $Q$ are parameters
proportional to the ADM mass and electric charge, respectively.  For
instance, in the spherical case and for zero cosmological constant one
has $m=\left(\frac{(d-2)\,\Sigma_{d-2}^{\,\,1}}{8\pi \, }\right)\,M$,
and $q^2=\left(\frac{(d-2)\,(d-3)}{2}\right)\,Q^2$, where $m$ and $q$
are the ADM mass and electric charge, respectively, and
$\Sigma_{d-2}^{\,\,1}$ is the area of the $(d-2)$-dimensional unit
sphere, $\Sigma_{d-2}^{\,\,1}= {2\pi^{\frac{d-
1}{2}}}/{\Gamma\left(\frac{d-1}{2}\right)}$.  When $\Lambda>0$ the ADM
quantities are not well defined. The zeros of $f(r)$ in
Eq.~(\ref{metric function}) give the gravitational radius $r_{\rm
g}$. The full metric vacuum solution given in Eqs.~(\ref{general
metric})-(\ref{metric function}), when extended to all $r$ ($0\leq
r<\infty$), yield black hole solutions.  Thus, in studying wormholes
the matter is put outside the gravitational radius $r_{\rm g}$.

We now consider two copies of the vacuum solution,
Eqs.~(\ref{general metric})-(\ref{angular metric differentials}),
removing from each copy the spacetime region given by
\begin{equation}
\Omega^{\pm} \equiv \{r^{\pm}\leq a\mid a > r_{\rm g}\}\,,
\end{equation}
where $a$ is a constant and $r_{\rm g}$ is the gravitational radius
given as the largest positive solution $r=r_{\rm g}$ to the equation
$f(r)=0$. The latter condition, $a > r_{\rm g}$, is important in order
not to have an event horizon. With the removal of these regions of
each spacetime, we are left with two geodesically incomplete
manifolds, with the following timelike hypersurfaces as boundaries
\begin{equation} \partial \Omega^{\pm}\equiv \{r^{\pm}=a\mid a> r_{\rm
g}\}\,.  \end{equation} The identification of these two timelike
hypersurfaces, $\partial \Omega^+ = \partial \Omega^- \equiv \partial
\Omega$, results in a manifold, now geodesically complete, where two
regions are connected by a wormhole. This wormhole has a throat at
$\partial \Omega$, which is the separating surface between the two
regions $\Omega^\pm$.
In fact, this cut-and-paste technique treats the throat
as a hypersurface between two regions of spacetime, where all the
exotic matter is concentrated, making the wormhole solution a
thin-shell wormhole solution. In this thin-shell case the location
of the two mouths coincide with that of the throat.
To continue the analysis we need to use the
Darmois-Israel formalism.
The intrinsic metric of this separating hypersurface $\partial \Omega$
can now be written as
\begin{equation}
ds^2_{\partial \Omega} = -d\tau^2+a^2(\tau)(d\Omega_{d-2}^k)^2\,,
\label{intrinsic metric}
\end{equation}
with $\tau$ being the proper time along the hypersurface $\partial
\Omega$, and $a(\tau)$, a quantity that defines the radius at which
the throat is located in each partial manifold $\Omega^{\pm}$, now a
function of the proper time of the throat.
In the bulk spacetime, the coordinates of this hypersurface are given
by $x^\alpha(\tau,\,\theta_1,\,\theta_2,\ldots,\theta_{d-2})$, with the
respective 4-velocity written as
\begin{equation}
u^\alpha_\pm \equiv \frac{dx^\alpha}{d\tau}\,.
\label{definition 4velocity}
\end{equation}
The intrinsic stress-energy tensor is defined through the Lanczos
equation as
\begin{equation}
S^i_j=-\frac{1}{8\pi}(\kappa_j^i-\delta_j^i\kappa^l_l)\,,
\label{intrinsic stress-energy tensor}
\end{equation}
where the indices written in the latin alphabet run as
$i=(\tau,\,\theta_1,\ldots,\,\theta_{d-2})$.
The quantity $\kappa_{ij}$ represent the
discontinuity in the extrinsic curvature $K_{ij}$, and one has
$\kappa_{ij}=K_{ij}^+-K_{ij}^-$. Each of the extrinsic curvatures, on
each of the original manifolds $\Omega^\pm$, is defined through
\begin{eqnarray}
K_{ij}^\pm &=& \frac{\partial x^\alpha}{\partial \xi^i}\frac{\partial
 x^\beta}{\partial \xi^j} \nabla^\pm_\alpha n_\beta\nonumber\\
         &=&-n_\gamma \left( \frac{\partial^2 x^\gamma}{\partial \xi^i
            \partial \xi^j}+\Gamma^{\gamma\,\pm}_{\alpha\beta} \frac{\partial
            x^\alpha}{\partial \xi^i}\frac{\partial
            x^\beta}{\partial\xi^j}\right)\,,
\label{extrinsic curvature}
\end{eqnarray}
where $n_\alpha$ is the unit normal to $\partial \Omega$ in the bulk
spacetime, the $\pm$ superscripts refer to the spacetime parcel
$\Omega^\pm$, and $\Gamma^{\gamma\,\pm}_{\alpha\beta}$ refers to the
respective spacetime Christoffell symbols.  The parametric equation
for the hypersurface $\partial \Omega$ can be written as
$f(x^\alpha(\xi^i))=0$. Using this equation we arrive at the formula for
the normal vector
\begin{equation}
n_\alpha = \pm \left| g^{\alpha\beta} \frac{\partial f}{\partial x^\alpha}
\frac{\partial f}{\partial x^\beta}\right|^{-\frac12} \frac{\partial
f}{\partial x^\alpha}\,.
\label{normal vector parametric}
\end{equation}
It can be ascertained that $n^\alpha n_\alpha=+1$, which makes it spacelike,
indeed $n^\alpha$ is normal vector to a timelike hypersurface.  Applying
Eqs.~(\ref{definition 4velocity}) and (\ref{normal vector parametric})
to our case, we have
$x^\alpha(\tau,\,\theta_1,\ldots,\,\theta_{d-2})=(t(\tau),
\,a(\tau),\,\theta_1,\ldots,\,\theta_{d-2})$, with the 4-velocity
written as
\begin{equation}
u^\alpha_\pm = \left(
 \frac{\sqrt{f(r)+\dot{a}^2}}{f(r)},\,\dot{a},\,0,\ldots,\,0\right)\,,
\label{4velocity}
\end{equation}
and the normal vector $n_{\alpha\pm}$ as
\begin{equation}
n_{\alpha\pm} =
\left(-\dot{a},\,\frac{\sqrt{f(r)+\dot{a}^2}}{f(r)},\,0,\ldots,\,0\right)\,,
\label{normal vector}
\end{equation}
where $\dot{}\equiv {\partial\;}/\partial\tau$.  This last result can
be arrived at through the relations $u^\alpha n_\alpha=0$ and $n^\alpha
n_\alpha=+1$, as well as through the relation (\ref{normal vector
parametric}).  Given the symmetry properties of the solutions we are
working with, the discontinuity of the extrinsic curvatures can be
written as $\kappa^i_j={\rm
diag}(\kappa^\tau_\tau,\,\kappa^{\theta_1}_{\theta_1},
\ldots,\,\kappa^{\theta_{d-2}}_{\theta_{d-2}})$.  This allows us to
write the surface intrinsic energy-momentum tensor as $S_j^i={\rm
diag}(-{\bf \sigma},\,\mathcal{P},\ldots,\,\mathcal{P})$, where ${\bf
\sigma}$ is the surface energy density and $\mathcal{P}$ is the
surface pressure. From the Lanczos equation, Eq.~(\ref{intrinsic
stress-energy tensor}), we obtain $ S_0^0 = -\frac{1}{8\pi}
\left(\kappa_0^0-\kappa_l^l\right) =
\frac{1}{8\pi}(\kappa^1_1+\,\kappa_2^2+\ldots+\,\kappa^{d-2}_{d-2})=
\frac{(d-2)}{4\pi\,a}\sqrt{f+\dot{a}^2}$, and so,
\begin{equation}
{\bf\sigma} = -\frac{(d-2)}{4\pi\,a}\sqrt{f+\dot{a}^2}\,.
\label{sigma}
\end{equation}
Using $S^1_1$ we obtain
\begin{equation}
\mathcal{P} = \frac{1}{8\pi} \frac{2\ddot{a} +
 f'}{\sqrt{f+\dot{a}^2}}-\frac{d-3}{d-2}{\bf\sigma}\,.
\label{pressure}
\end{equation}
For the latter results we have used the following expressions for the
extrinsic curvatures, defined in Eq.~(\ref{extrinsic curvature}),
$ K_0^{0\,\pm} = \pm \frac{\frac12 f'+\ddot{a}}{\sqrt{f+\dot{a}^2}}$,
$K_1^{1\,\pm} = \pm \frac1a \sqrt{f+\dot{a}^2}$, $K_2^{2\,\pm} =
\pm \frac1a \sqrt{f+\dot{a}^2}$\label{extrinsic curvatures for omega},
and so on. One can substitute one of the two equations
(\ref{sigma}) and (\ref{pressure}) by
the conservation equation, which can be written as
\begin{equation}
\frac{d}{d\tau}({\bf
\sigma}\,a^{d-2})+\mathcal{P}\frac{d}{d\tau}(a^{d-2})=0\,.
\label{conservation equation}
\end{equation}
In order to solve Eqs.~(\ref{sigma})-(\ref{pressure}),
or if we prefer Eqs.~(\ref{sigma}) and (\ref{conservation equation}),
for ${\bf \sigma}(\tau)$ and $a(\tau)$, one would
need to choose an equation
of state, the most simple one would be a cold equation of state
$\mathcal{P}=\mathcal{P}({\bf\sigma})$.

\subsection{Static wormholes}
\label{static wormholes}

Now, we resort to a static shell for which
$\dot{a}=\ddot{a}=0$. In
this case Eqs.~(\ref{sigma}) and (\ref{pressure}) reduce to
\begin{eqnarray}
\density &=& -\frac{(d-2)}{4\pi\,a}\sqrt{f}\,,
\label{static sigma}\\
\pressure &=& \frac{a\,f'+2(d-3)\,f}{8\pi\,a\sqrt{f}}\,,
\label{static pressure}
\end{eqnarray}
A general equation of state of the form
$\pressure=\pressure(\density)$ turns
the terms in $f(a)$, such as $M$, $Q$, $a$, and $\Lambda$ into related
terms, that is, there is a relation between these terms, making
them dependent upon each other.

\section{Wormhole properties: Energy conditions
and test particle motion}
\label{ex}

\subsection{Wormhole and domain walls}

These traversable wormhole solutions are a natural extension of the
corresponding black hole solutions upon the addition of exotic
matter. For the plane and hyperbolic traversable wormhole solutions,
infinity carries the same topology as the the topology of the $t={\rm
constant}$, $r={\rm constant}$ $(d-2)$-dimensional surfaces of the
background spacetime. Thus, the construction of these wormholes does
not alter the topology of the background spacetime (i.e., spacetime is
not multiply-connected).  Therefore, these solutions can instead be
considered higher-dimensional domain walls or branes connecting two
universes, and as such, in general, do not allow time travel.

\subsection{Energy conditions}

\subsubsection{The throat shell}
Now we turn to the issue of the energy conditions on the shell.
In our case the weak energy
condition is given by $\density+\pressure\geq0$ and
$\density\geq0$. Since, in the reference frame we
are working, the surface energy density is
negative, see Eq.~\eq{static sigma},  this
means that the weak energy condition is violated as usual
for wormholes.
The null energy condition requires only that $\density+\pressure\geq0$.
Using \eq{static sigma}-\eq{static pressure} this
implies the following inequality
\begin{eqnarray}
a^{d-3}\left(2\,a^{d-3}k-(d-1)M\right)+2(d-2)Q^2&\geq&0\,.
\label{nec for our case}
\end{eqnarray}
We can further require that the strong
energy condition holds, i.e., $\density+(d-2)\pressure\geq0$, which
using \eq{static sigma}-\eq{static pressure} yields
\begin{eqnarray}
(d-4)k-\frac{(d-3)}{3}\Lambda\,a^2-\frac{(d-5)M}{2\,a^{d-3}}-
\frac{Q^2}{a^{2(d-3)}}\geq0\,.
\label{sec for our case}
\end{eqnarray}

\subsubsection{The two exterior regions to the shell}
Now we turn to the issue of the energy conditions outside the shell,
i.e., for $r>a$.  Here only the electromagnetic field and $\Lambda$
contribute to the energy-momentum tensor of the Einstein equations,
see \cite{rahamanetal} for the case of spherical symmetry $(k=1)$.
The energy-momentum tensor of the electromagnetic field is
$
T^{\rm \rm em}_{\alpha\beta} = \frac{1}{4\pi}\left(-F^{\gamma}_\alpha
 F_{\beta\gamma} - \frac14 g_{\alpha\beta}
 F_{\gamma\delta}F^{\gamma\delta}\right)
\label{emtensorforelc}
$.
The only nonzero component of the electromagnetic field is
$
E_r = \frac{Q}{r^{d-2}}
\label{er}
$.
Calculating the $T_{00}^{\rm em}$ component of the energy-momentum tensor
yields
$
T^{\rm em}_{00} = \frac{1}{8\pi}\,f(r)\,\frac{Q^2}{r^{2(d-2)}}
\label{t00}
$.
We can write the $T_{11}^{\rm em}$ component as
$
T^{\rm em}_{11} = - \frac{1}{8\pi}\,f(r)^{-1}\,\frac{Q^2}{r^{2(d-2)}}
\label{t11}
$.
We have also to include the vacuum energy represented by the
cosmological constant. The corresponding
vacuum energy-momentum tensor
can be written as
$
T_{\alpha\beta}^{\rm vac}=-\frac{\Lambda}{8\pi} g_{\alpha\beta}
\label{t_vac}
$.
Thus,
$
T_{00}^{\rm vac} = \frac{\Lambda}{8\pi} f(r)
\label{t00_vac}
$,
$
T_{11}^{\rm vac} = -\frac{\Lambda}{8\pi} f^{-1}(r)\
\label{t11_vac}
$,
$
T_{22}^{\rm vac} = -\frac{\Lambda}{8\pi} r^2
\label{t22_vac}
$, and so on.
Using $u_\alpha=(-\sqrt{f},\,0,\ldots,\,0)$ for a timelike, future directed vector field, 
and 
$
T_{00} = \rho f(r)\
\label{emtensorperfectfluid00}
$,
$
T_{11} = p_r f(r)^{-1} \label{emtensorperfectfluid11}
$,
$
T_{22} = p_{\theta_1} r^2 \label{emtensorperfectfluid22}
$,
$
T_{33}^{k=1} = p_{\theta_2} r^2 \sin\theta_1^2
\label{emtensorperfectfluid33k1}
$,
$
T_{33}^{k=0} = p_{\theta_2} r^2 \label{emtensorperfectfluid33k0}
$,
$
T_{33}^{k=-1} = p_{\theta_2} r^2
\sinh\theta_1^2 \label{emtensorperfectfluid33k-1}
$, we can establish that the weak energy condition, $T_{\mu\nu}u^\mu u^\nu \geq 0$, 
is satisfied if
\begin{eqnarray}
Q^2 &\geq& - \Lambda \, r^{2(d-2)}\,,
\label{wec1}
\end{eqnarray}
implying that if $\Lambda<0$, then the weak energy condition is satisfied only if 
$\left|Q^2\right|\geq \sqrt{\left|\Lambda \right|}\,r^{d-2}$, 
and if $\Lambda<0$, then the weak energy condition is always satisfied, because 
$Q^2>0$ and $\Lambda>0$ make it always true. 
There is no dependence on the topological factor $k$.
So, outside the shell, where the contributions to the
energy-momentum tensor come from the electromagnetic field and the
cosmological constant, the weak energy condition is not satisfied
except in a finite region, if $\left|Q\right|$ is large enough, in the 
case of anti-de Sitter spacetimes.
As to the null energy condition, it always holds. This can be gleaned from 
the fact that $T_{\mu\nu} k^\mu k^\nu = 0$, where the $k^\mu$ are null vectors,
defined as $k^{\mu}=(-\frac{1}{\sqrt{f}},\sqrt{f},0,\ldots,0)$, where $f$ is again
the metric function, and the $T_{\mu\nu}$ is the same as above, 
the sum of the electromagnetic and vacuum energy-momentum tensors. 

\subsection{Attraction and repulsion of the wormhole on
test particles}

In order to complete the general considerations on the properties of
the static wormholes of the family of solutions under study, we
address the issue of the attractive or repulsive character of the
traversable wormhole on test particles.  First, let us recall the
4-velocity in Eq.~\eq{4velocity} now for a test particle. The
expression contains the term $\dot{a}^2$ inside the square root.  For
the static wormhole case under consideration it holds that
$\dot{a}^2=0$. So the 4-velocity is now written as
\begin{equation}
u^\alpha_\pm = \left(\frac{1}{\sqrt{f(r)}},\,0,\ldots,\,0\right)
\label{static 4velocity}
\end{equation}
The 4-acceleration is $a^\alpha=u^\alpha_{;\beta}\,u^\beta$. Now,
given the expression for $u^\alpha$ in Eq.~\eq{static 4velocity} we
have
\begin{equation}
a^\alpha={u^\alpha}_{;0}\,u^0={u^\alpha}_{;0} \frac{1}{\sqrt{f(r)}}\,.
\label{4acceleration}
\end{equation}
We can show that
$
{u^\alpha}_{;0} = {u^\alpha}_{,0} +
\Gamma_{\alpha 0}^{\alpha}\,u^{\alpha} =
\Gamma_{00}^{\alpha}u^0
\label{actual 4velocity}
$.
The only nonzero component of this is $\alpha=1$, such that
$u^1_{;0}=\Gamma^1_{00}u^0$. From this we know that
$a^\alpha=(0,\,a^r,\,0,\ldots,\,0)$, where $a^r=\frac12
f'(r)$. Expanding, we get
\begin{equation}
a^r = -\frac{\Lambda}{3}
r+\frac{(d-3)M}{2\,r^{d-2}}-\frac{(d-3)Q^2}{r^{2d-5}}\,.
\label{actual 4acceleration}
\end{equation}
The geodesic equation,
$
\frac{d^2 r}{d\tau^2} + \Gamma^1_{00}\dot{x}^0\dot{x}^0=0
\label{geodesic equation}
$,
allows us to write
\begin{equation}
\frac{d^2 r}{d\tau^2} = -\Gamma^1_{00}
\left(\frac{d^2t}{d\tau^2}\right)=-a^r\,.
\label{radial acceleration}
\end{equation}
As in \cite{lemoslobo} an observer must maintain a proper acceleration
with the radial component given in Eq.~\eq{actual 4acceleration} in
order to remain at rest. Again, the wormhole is attractive if $a^r>0$
and repulsive if $a^r<0$, which depends on the balancing of the
parameters in Eq.~\eq{actual 4acceleration}.

\section{Stability analysis: linear stabilty}
\label{linearized stability}

\subsection{General considerations: the stability equations and criteria}
\label{general considerations}


The equations of motion (\ref{sigma}) and (\ref{conservation equation})
can be put in the more useful form
\begin{equation}
\dot{a}^2  - \left\{-\frac{\Lambda}{3}a^2-\frac{M}{a^{d-3}}+
\frac{Q^2}{a^{2(d-3)}}-\frac{16\,\pi^2\,a^2}{(d-2)^2}
\density^2(a)\right\}
=-k\,,
\label{a square equation}
\end{equation}
and
\begin{equation}
\dot{{\bf \sigma}}=-(d-2)\frac{\dot{a}}{a}({\bf \sigma}+\mathcal{P})\,,
\label{conservation equation2}
\end{equation}
respectively. One can choose a particular equation of state, for instance, an
equation of state taken from the
generic cold  equation of state
${\cal P}={\cal P}(\sigma)$. Then we can integrate
Eq.~(\ref{conservation equation2}) to yield
the general solution
$
\ln{a}=-\frac{1}{d-2}\,\int
\frac{d\density}{\density+\pressure(\density)}
$,
which, if we wish,
can be formally inverted to provide the function
$\density=\density(a)$, i.e., the wormhole surface density as
a function of its own radius. Substituting this equation
for $\ln a$
into Eq.~(\ref{a square equation}) determines the the motion
of the throat, and so its stability.

We follow other path provided by Poisson and Visser
\cite{poisson}. In order to test stability we consider a linear
perturbation around those static solutions found in
Sec.~\ref{static wormholes}.  Let us take $a_0$ as
the radius of the static solution. The respective static values of the
surface energy density and the surface pressure are given by
Eqs.~(\ref{static sigma})-(\ref{static pressure}).  To know whether
the equilibrium solution is stable or not, one must analyze the
shell's equation of motion near the equilibrium solution.  Following
\cite{poisson} we put quite generally
\begin{equation}
{\cal P}={\cal P}(\sigma)\,,
\label{cold}
\end{equation}
i.e., we impose a generic, as opposed to a particular, cold equation
of state.  Now, Eq.~(\ref{a square equation}) can be written in the
more useful form
\begin{equation}
\dot{a}^2=-V(a)\,,
\label{a dot and V}
\end{equation}
with $V(a)$ given by
\begin{equation}
V(a) = k-\frac{\Lambda}{3}a^2-\frac{M}{a^{d-3}}+
\frac{Q^2}{a^{2(d-3)}}-\frac{16\,\pi^2\,a^2}{(d-2)^2}\density(a)^2\,,
\label{va}
\end{equation}
or more compactly as,
\begin{equation}
V(a) = f(a)-\frac{16\,\pi^2\,a^2}{(d-2)^2}\density(a)^2\,.
\label{va compact}
\end{equation}
In the study of the stability of a static solution of radius $a_0$, of
course we are considering $f(a_0)$ in the interval $f(a_0)>0$ so that
no event horizon is present. Our initial condition before the cutting
and gluing operation is that $a>r_{\rm g}$, hence $a_0>r_{\rm g}$.
A linearization is going to be done in order to determine whether and
in what conditions the static solutions with a radius $a_0$ are stable
under a linear perturbation around $a_0$.  The values of the density
and pressure are given in Eqs.~\eq{static sigma} and \eq{static
pressure} for the case of a static solution for the throat.  For the
linearization, we make a Taylor expansion of the function $V(a)$
around the static radius $a_0$,
\begin{equation}
V(a) = V(a_0) + V'(a_0)(a-a_0) + \frac12 V''(a_0)(a-a_0)^2 + O[(a-a_0)^3]\,,
\label{taylor va}
\end{equation}
with a prime $'$ corresponding to a derivative with respect to $a$.
By definition, if we are dealing with a static configuration then
$V(a_0)=0$, because of Eq.~\eq{a dot and V}. One can also
show that $\ddot{a}=-\frac12 V'(a)$. With this relation, as a
static configuration also demands that $\ddot{a}=0$, we also have
$V'(a_0)=0$, where $V'(a)=\dot{V}/\dot{a}$.  From
Eq.~\eq{cold} one can define a quantity $\eta$ as,
\begin{equation}
\etab(\density) \equiv
\frac{d\pressure}{d\density}\,,
\label{eta sigma}
\end{equation}
or, when preferable, $\etab(\density) =\frac{\pressure'}{\density'}$.
With this definition the second derivative $V''(a)$ can be written as,
\begin{eqnarray}
V''(a) &=& -\frac{2\Lambda}{3} - \frac{(d-2)(d-3)M}{a^{d-1}} +
\frac{2(d-3)(2d-5)Q^2}{a^{2(d-2)}}\nonumber\\
      &{}&
      -\frac{32\pi^2}{(d-2)^2}\left\{[(d-3)\density+(d-2)\pressure]^2
      +(d-2)\density(\density+\pressure)(d-3+(d-2)\etab)\right\}\,.
\label{vii}
\end{eqnarray}
From Eqs.~\eq{a dot and V} and \eq{taylor va},
the equation of motion of the wormhole throat is given by
\begin{equation}
\dot{a}^2 = -\frac12 V''(a_0)(a-a_0)^2+O\left[(a-a_0)^3\right]\,.
\label{linear eom}
\end{equation}
In order to insure stability we have to guarantee that the second
derivative evaluated at the radius of the static configuration $a_0$
is positive, i. e., $V''(a_0)>0$. This is going to present us with
conditions on $\etab(\density)$.  To write the conditions in a
manageable form we define $f_0\equiv f(a_0)$ and
$\etab_0\equiv\etab(\density_0)$, for the quantities considered at
$a_0$. A prime on $f$ indicates derivation with respect to
$a$. The conditions can be put thus
\begin{eqnarray}
\etab_0 <
\frac{a_0^2\left(f_0'\right)^2-2\,a_0^2\,f_0''\,f_0}{2(d-2)\,f_0
\left(-2k+\frac{(d-1)M}{a_0^{d-3}}-\frac{2(d-2)Q^2}{a_0^{2(d-3)}}\right)}
-\frac{d-3}{d-2}\,\,\,\,\,\, &{\rm if}& \,\,\,
 -2k+\frac{(d-1)M}{a_0^{d-3}}-\frac{2(d-2)Q^2}{a_0^{2(d-3)}}<0\,,
\label{conditions for etab1}\\
\etab_0=-\infty\,, {\rm or}\,\, \etab_0=+\infty\,\,,
{\rm depending}\, {\rm on}\,
{\rm the}\, {\rm branching}\,
\,\,\,\,\,\,\,\,\, &{\rm if}& \,\,\,
-2k+\frac{(d-1)M}{a_0^{d-3}}-\frac{2(d-2)Q^2}{a_0^{2(d-3)}}=0\,,
\label{conditions for etab2}\\
\etab_0 >
\frac{a_0^2\left(f_0'\right)^2-2\,a_0^2\,f_0''\,f_0}{2(d-2)\,f_0
\left(-2k+\frac{(d-1)M}{a_0^{d-3}}-\frac{2(d-2)Q^2}{a_0^{2(d-3)}}\right)}
-\frac{d-3}{d-2}\,\,\,\,\,\, &{\rm if}& \,\,\,
 -2k+\frac{(d-1)M}{a_0^{d-3}}-\frac{2(d-2)Q^2}{a_0^{2(d-3)}}>0\,.
\label{conditions for etab3}
\end{eqnarray}
The quantity $f_0$ is given by Eq.~\eq{metric function} evaluated at
$a_0$. The quantities $f_0'$ and $f_0''$ are given by
\begin{eqnarray}
f_0'&=& -\frac{2\Lambda a_0}{3} + \frac{(d-3)M}{a_0^{d-2}} -
\frac{2(d-3)Q^2}{a_0^{2d-5}}\,,
\label{f0 prime}\\
f_0''&=& -\frac{2\Lambda}{3} - \frac{(d-3)(d-2)M}{a_0^{d-1}} +
\frac{2(d-3)(2d-5)Q^2}{a_0^{2d-4}}\,,
\label{f0 primeprime}
\end{eqnarray}
evaluated at $a_0$ respectively.
Eqs.~\eq{conditions for etab1}-\eq{conditions for etab3},
together with
Eqs.~\eq{f0 prime}-\eq{f0 primeprime},
generalize
what has been found in previous papers.

It is interesting to see to what does this lead in the case that the
cold equation of state ${\cal P}={\cal P}(\sigma)$ in Eq.~\eq{cold},
assumes a particular form, $\pressure=\omega\density$, with
$\omega<0$. This is a dark energy equation of state.  The expression
for the second derivative $V''(a_0)$ is now $ V''(a_0) = f_0'' -
\frac{32\pi^2\density^2}{(d-2)^2}
\left\{[(d-3)+(d-2)\,\omega]^2-(d-2)(1+\omega)[(d-3)+(d-2)\,\omega]
\right\} $, so that the condition $V''(a_0)>0$ implies $
\frac{f_0''\,(d-2)^2}{32\pi^2\density^2}>
\left\{[(d-3)+(d-2)\,\omega]^2-(d-2)(1+\omega)[(d-3)+(d-2)\,\omega]
\right\} $.  This is not conclusive in general, but contains as
particular cases the results found in previous papers, e.g.,
\cite{lemoslobo}.

\subsection{Particular cases studied in the literature and a new example}

It is now interesting to reduce our general results to some well known
cases already studied for particular choices of the parameters $d$,
$k$, $\Lambda$, and $Q$. We also give a new example.

\subsubsection{Poisson-Visser, $d=4$, $k=1$, $\Lambda=0$, and $Q=0$}
Poisson and Visser \cite{poisson} were the first to study the linear
stability of wormholes. Perturbations were done around some
four-dimensional static wormhole solution with spherical symmetry with
no cosmological constant and no charge.
Putting $d=4$, $k=1$, $\Lambda=0$, $Q=0$,
calling $m$ the ADM mass, so that for this case our mass
parameter $M$ is $M\to 2m$,
our
formulas \eq{conditions for etab1}-\eq{conditions for etab3}
yield
\begin{eqnarray}
\etab_0 <
-\frac{1-3m/a_0+3(m/a_0)^2}
{2(1-2m/a_0)(1-3m/a_0)}\,\,\,\,\,\, &{\rm if}& \,\,\,
-1+\frac{3m}{a_0} <0\,,
\label{conditions for etab1 Poisson Visser}\\
\etab_0=-\infty\,, {\rm or}\,\, \etab_0=+\infty\,\,,
{\rm depending}\, {\rm on}\,
{\rm the}\, {\rm branching}\,
\,\,\,\,\,\,\,\,\, &{\rm if}& \,\,\,
-1+\frac{3m}{a_0} =0\,,
\label{conditions for etab2 Poisson Visser}\\
\etab_0 >
-\frac{1-3m/a_0+3(m/a_0)^2}
{2(1-2m/a_0)(1-3m/a_0)}\,\,\,\,\,\, &{\rm if}& \,\,\,
-1+\frac{3m}{a_0} >0\,.
\label{conditions for etab3 Poisson Visser}
\end{eqnarray}
This is precisely what Poisson and Visser \cite{poisson} obtained.

\subsubsection{Lobo-Crawford, $d=4$, $k=1$, $\Lambda\neq0$, and $Q=0$}
Lobo and Crawford \cite{lobolinear} extended the
Poisson and Visser study by considering
$\Lambda\neq0$ in the analysis of linear
stability of wormholes.
Then putting again $M\to 2m$,
and  $d=4$, $k=1$, $\Lambda$ generic, and $Q=0$ in our
formulas \eq{conditions for etab1}-\eq{conditions for etab3}, one
finds
\begin{eqnarray}
\etab_0 <-\frac{1- \frac{3m}{a_0}+
 \frac{3m^2}{a_0^2}-\Lambda m
 a_0}{2\left(1-\frac{2m}{a_0}-\frac13\Lambda a_0^2\right)
\left(1-\frac{3m}{a_0}\right)}
\,\,\,\,\,\, &{\rm if}& \,\,\,
-1+\frac{3m}{a_0} <0\,,
\label{conditions for etab1 Lobo Crawford}\\
\etab_0=-\infty\,, {\rm or}\,\, \etab_0=+\infty\,\,,
{\rm depending}\, {\rm on}\,
{\rm the}\, {\rm branching}\,
\,\,\,\,\,\,\,\,\, &{\rm if}& \,\,\,
-1+\frac{3m}{a_0} =0\,,
\label{conditions for etab2 Lobo Crawford}\\
\etab_0 >-\frac{1- \frac{3m}{a_0}+
 \frac{3m^2}{a_0^2}-\Lambda m
 a_0}{2\left(1-\frac{2m}{a_0}-\frac13\Lambda a_0^2\right)
\left(1-\frac{3m}{a_0}\right)}
\,\,\,\,\,\, &{\rm if}& \,\,\,
-1+\frac{3m}{a_0} >0\,.
\label{conditions for etab3 Lobo Crawford}
\end{eqnarray}
This is what Lobo and Crawford \cite{lobolinear}
obtained. Putting further $\Lambda=0$ in
Eqs.~\eq{conditions for etab1 Lobo Crawford}-\noindent
\eq{conditions for etab3 Lobo Crawford} one obtains
Eqs.~\eq{conditions for etab1 Poisson Visser}-\noindent
\eq{conditions for etab3 Poisson Visser}.

\subsubsection{Eiroa-Romero, $d=4$, $k=1$, $\Lambda=0$, and $Q\neq0$}

Eiroa and Romero \cite{eiroa}
discussed for the first time wormhole stability for systems with
electric charge $Q\neq0$.
Putting $M=2m$, with $m$ being the ADM mass of the
$d=4$ solution, $Q=q$, with $q$ being the ADM charge of the
$d=4$ solution, $k=1$, $\Lambda=0$, we recover the
Reissner-Nordstr\"om
wormhole system and stability given
in \cite{eiroa}.
The relevant condition, i.e.,
the condition on the right hand side of
Eqs.~\eq{conditions for etab1}-\eq{conditions for etab3},
yields that either $a_0$ is larger or smaller than
$3 m+\sqrt{3 m^2-2q^2}$, which is the same
result as of \cite{eiroa}. The no gravitational radius
condition, $a_0>r_{\rm g}$, implies
$
a_0 > m + \sqrt{m-q^2}
\label{above horizon eiroa romero}
$.
In this case, our
formulas \eq{conditions for etab1}-\eq{conditions for etab3}
give for the case of $\frac{|q|}{2m}<1$,
\begin{eqnarray}
\etab_0 <-
\frac{\left(1-\frac{m}{a_0}\right)^3+
\frac{m}{a_0^3}\left(m^2-q^2\right)}
{2\left(1-\frac{2m}{a_0}+\frac{q^2}{a_0}\right)\left(1-\frac{3m}{a_0}+
\frac{2q^2}{a_0^2}\right)}
\,\,\,\,\,\, &{\rm if}& \,\,\,
-1+\frac{3m}{a_0}-\frac{2q^2}{a_0^2} <0\,,
\label{conditions for etab1 Eiroa Romero}\\
\etab_0=-\infty,\, {\rm or}\,\, \etab_0=+\infty,\,
{\rm depending}\, {\rm on}\,
{\rm the}\, {\rm branching}\,
\,\,\,\,\,\,\,\,\, &{\rm if}& \,\,\,
-1+\frac{3m}{a_0}-\frac{2q^2}{a_0^2} =0\,,
\label{conditions for etab2 Eiroa Romero}\\
\etab_0 >-
\frac{\left(1-\frac{m}{a_0}\right)^3+
\frac{m}{a_0^3}\left(m^2-q^2\right)}
{2\left(1-\frac{2m}{a_0}+\frac{q^2}{a_0}\right)\left(1-\frac{3m}{a_0}+
\frac{2q^2}{a_0^2}\right)}
\,\,\,\,\,\, &{\rm if}& \,\,\,
-1+\frac{3m}{a_0}-\frac{2q^2}{a_0^2}  >0\,.
\label{conditions for etab3 Eiroa Romero}
\end{eqnarray}
This is what Eiroa and Romero \cite{eiroa} obtained.  Putting further
$q=0$ in Eqs.~\eq{conditions for etab1 Eiroa Romero}-\noindent
\eq{conditions for etab3 Eiroa Romero} one obtains Eqs.~\eq{conditions
for etab1 Poisson Visser}-\noindent \eq{conditions for etab3 Poisson
Visser}. One deduces from Eqs.~\eq{conditions for etab1 Eiroa
Romero}-\noindent \eq{conditions for etab3 Eiroa Romero} that electric
charge tends to destabilize the system. We have presented the case
$\frac{|q|}{2m}<1$ for way of comparison, for the other cases and a
detailed analysis see \cite{eiroa}.

\subsubsection{Lemos-Lobo, $d=4$, $k=0$, $\Lambda\neq0$, and $Q=0$}

Lemos and Lobo \cite{lemoslobo} studied the stability of
four-dimensional planar wormholes, i.e., wormholes
with $k=0$.  Putting $d=4$, $k=0$, $\Lambda\neq0$, and $Q=0$, and
noting that for the planar case $M$ can be consider the ADM mass
$m$ \cite{lemosbhs}, $M=m$, our formulas \eq{conditions for
etab1}-\eq{conditions for etab3} give
\begin{eqnarray}
\etab_0 <-
\frac{\frac{m}{2 a_0}-\frac\Lambda3 a_0^2}{2\left(\frac{m}{ a_0}
+\frac\Lambda3
   a_0^2\right)}
\,\,\,\,\,\, &{\rm if}& \,\,\,
\frac{3m}{a_0} <0\,,
\label{conditions for etab1 Lemos Lobo}\\
\etab_0=-\infty,\, {\rm or}\,\, \etab_0=+\infty,\,
{\rm depending}\, {\rm on}\,
{\rm the}\, {\rm branching}\,
\,\,\,\,\,\,\,\,\, &{\rm if}& \,\,\,
\frac{3m}{a_0} =0\,,
\label{conditions for etab2 Lemos Lobo}\\
\etab_0 >-
\frac{\frac{m}{2 a_0}-\frac\Lambda3 a_0^2}{2\left(\frac{m}{ a_0}
+\frac\Lambda3
   a_0^2\right)}
\,\,\,\,\,\, &{\rm if}& \,\,\,
\frac{3m}{a_0}  >0\,.
\label{conditions for etab3 Lemos Lobo}
\end{eqnarray}
This is what Lemos and Lobo obtained \cite{lemoslobo}.
In fact, Lemos and Lobo worked out the case
with negative cosmological constant
and have defined $\alpha$ such that $\alpha^2=-\frac{\Lambda}{3}$.
One does not need to make such a restriction and can consider
wormholes valid for $\Lambda<0$ as well as $\Lambda\geq0$ as we show here.
The mass $m$ here and the mass in \cite{lemoslobo} are different,
$m_{\rm here}=m_{\rm LL}/\alpha$. The case $m=0$ is the case of
no wormhole, where spacetime is just de Sitter, flat, or anti-de Sitter,
depending on $\Lambda$.

\subsubsection{Rahaman-Kalam-Chakraborty, $d=d$, $k=1$,
$\Lambda=0$, and $Q\neq0$}

Another particular case seen in the literature
was studied by Rahaman-Kalam-Chakraborty
\cite{rahamanetal}, where $d$ is left general, the spherical $k=1$ geometry
is chosen, the cosmological constant is set to zero $\Lambda=0$, and
there is a nonzero electric charge $Q\neq0$. Their main expression for
$\etab_0$ is taken from our general expression in Eqs.~\eq{conditions
for etab1} and \eq{conditions for etab2}, with the appropriate
replacements.
Putting $d=d$, $k=1$, $\Lambda=0$, and $Q\neq0$,
our formulas \eq{conditions for
etab1}-\eq{conditions for etab3} give
\begin{eqnarray}
\etab_0 <
\frac{a_0^2\left(f_0'\right)^2-2\,a_0^2\,f_0''\,f_0}{2(d-2)\,f_0
\left(-2+\frac{(d-1)M}{a_0^{d-3}}-\frac{2(d-2)Q^2}{a_0^{2(d-3)}}\right)}
-\frac{d-3}{d-2}\,\,\,\,\,\, &{\rm if}& \,\,\,
 -2+\frac{(d-1)M}{a_0^{d-3}}-\frac{2(d-2)Q^2}{a_0^{2(d-3)}}<0\,,
\label{conditions for etab1 Rahaman Kalam Chakraborty}\\
\etab_0=-\infty,\, {\rm or}\,\, \etab_0=+\infty,\,
{\rm depending}\, {\rm on}\,
{\rm the}\, {\rm branching}\,
\,\,\,\,\,\,\,\,\, &{\rm if}& \,\,\,
-2+\frac{(d-1)M}{a_0^{d-3}}-\frac{2(d-2)Q^2}{a_0^{2(d-3)}}=0\,,
\label{conditions for etab2 Rahaman Kalam Chakraborty}\\
\etab_0 >
\frac{a_0^2\left(f_0'\right)^2-2\,a_0^2\,f_0''\,f_0}{2(d-2)\,f_0
\left(-2+\frac{(d-1)M}{a_0^{d-3}}-\frac{2(d-2)Q^2}{a_0^{2(d-3)}}\right)}
-\frac{d-3}{d-2}\,\,\,\,\,\, &{\rm if}& \,\,\,
-2+\frac{(d-1)M}{a_0^{d-3}}-\frac{2(d-2)Q^2}{a_0^{2(d-3)}}>0\,.
\label{conditions for etab3 Rahaman Kalam Chakraborty}
\end{eqnarray}
Eqs.~\eq{conditions for etab1 Rahaman Kalam Chakraborty}-\noindent
\eq{conditions for etab3 Rahaman Kalam Chakraborty}
complete the stability conditions given in
\cite{rahamanetal}, in \cite{rahamanetal} only
Eq.~\eq{conditions for etab1 Rahaman Kalam Chakraborty} is given.  In
the inequalities \eq{conditions for etab1 Rahaman Kalam
Chakraborty}-\noindent \eq{conditions for etab3 Rahaman Kalam
Chakraborty}\noindent, one can explicitly write the expressions for
$f_0$, $f'_0$, and $f''_0$ in terms of the ADM mass $m$, the ADM
charge $q$, and the other quantities.  In this case, since the
geometry is spherical, $k=1$, the relation between the mass parameter
$M$ and the ADM mass $m$ is given by
\begin{eqnarray}
M= \frac{16\,\pi\,\Gamma\left(\frac{d-1}{2}\right)}{(d-2)\,2\pi^{\frac{d-
1}{2}}}\;
m \,,
\label{spherical adm mass 2}
\end{eqnarray}
where $\Gamma(z)$ is the Gamma Function, and between
the charge parameter $Q$
and the ADM charge $q$ is given by
\begin{eqnarray}
Q^2=\frac{2}{(d-2)(d-3)}\;q^2\,.
\label{spherical adm charge 2}
\end{eqnarray}
Thus, the inequality
\eq{conditions for etab1 Rahaman Kalam Chakraborty},
can be written as
\begin{eqnarray}
\eta_0 &<&
-\frac{d-3}{d-2}\,\frac{
1
-\frac32\frac{16\,\pi\,\Gamma\left(\frac{d-1}{2}\right)}{(d-2)
\,2\pi^{\frac{d-
 1}{2}}}\frac{m}{a_0^{d-3}}
-
\frac{
\left(\frac{16\,\pi\,\Gamma\left(\frac{d-1}{2}\right)}{(d-2)
\,2\pi^{\frac{d-
1}{2}}}\;m\right)
\left(\frac{2}{(d-2)(d-3)}\;q^2m\right)}{2a_0^{3(d-3)}}
+\frac{(d-1)\left(\frac{16\,\pi\,\Gamma
\left(\frac{d-1}{2}\right)}{(d-2)\,
2\pi^{\frac{d-1}{2}}}\;
m\right)^2-4(d-4)\left(\frac{2}{(d-2)(d-3)}\;q^2
\right)}{4a_0^{2(d-3)}}
}
{
\left(1-\frac{\frac{16\,\pi\,\Gamma
\left(\frac{d-1}{2}\right)}{(d-2)
\,2\pi^{\frac{d-
1}{2}}}\;
m}{a_0^{d-3}}+\frac{\frac{2}{(d-2)(d-3)}\;
q^2}{a_0^{2(d-3)}}\right)
\left(1-\frac{(d-1)\frac{16\,\pi\,\Gamma\left(\frac{d-1}{2}
\right)}{(d-2)\,2\pi^{\frac{d-
1}{2}}}\;
m}{2a_0^{d-3}}+\frac{(d-3)\frac{2}{(d-2)(d-3)}\;
q^2}{a_0^{2(d-3)}}\right)
}\,,\nonumber\\
&& \textrm{if}\,\,\,
-2+\frac{(d-1)\frac{16\,\pi\,\Gamma\left(\frac{d-1}{2}
\right)}{(d-2)\,2\pi^{\frac{d-
1}{2}}}\;
m}{a_0^{d-3}}-\frac{2(d-2)\frac{2}{(d-2)(d-3)}\;
q^2}{a_0^{2(d-3)}}<0\,.
\label{etab for Rahaman Kalam Chakraborty 2}
\end{eqnarray}
To rewrite the
conditions \eq{conditions for etab2 Rahaman Kalam Chakraborty} and
\eq{conditions for etab3 Rahaman Kalam Chakraborty}
one has only to make the appropriate changes in the
inequality sign.

\subsubsection{A new example in several dimensions $d=4\,,5\,,
\infty$,$\;$ $k=1,0,-1$,$\;$ $\Lambda\neq0$, $\;$and $Q=0$}

\begin{figure}[htb]
\begin{center}
\includegraphics[width=0.25\textwidth]{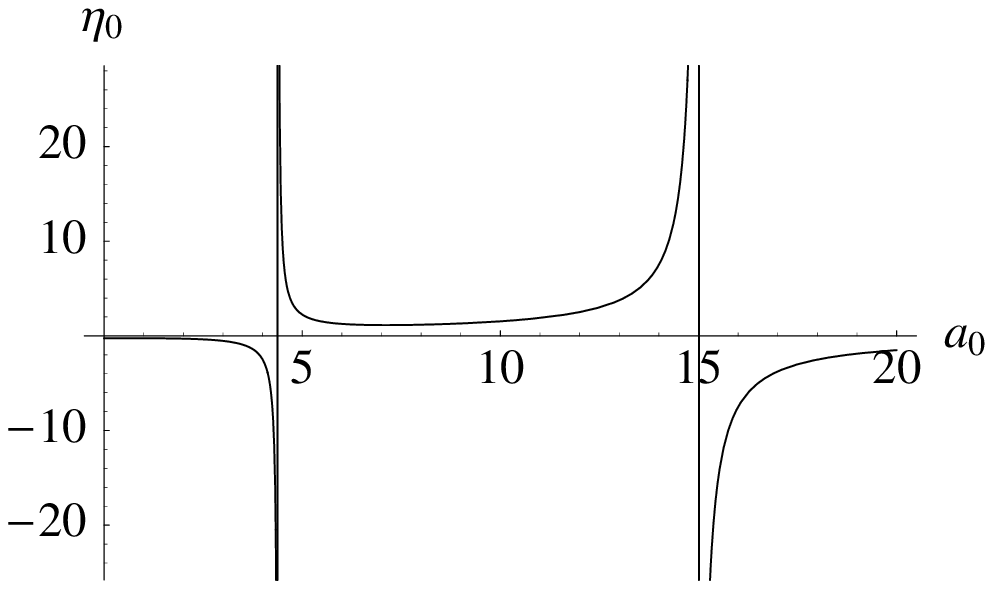} \quad
\includegraphics[width=0.25\textwidth]{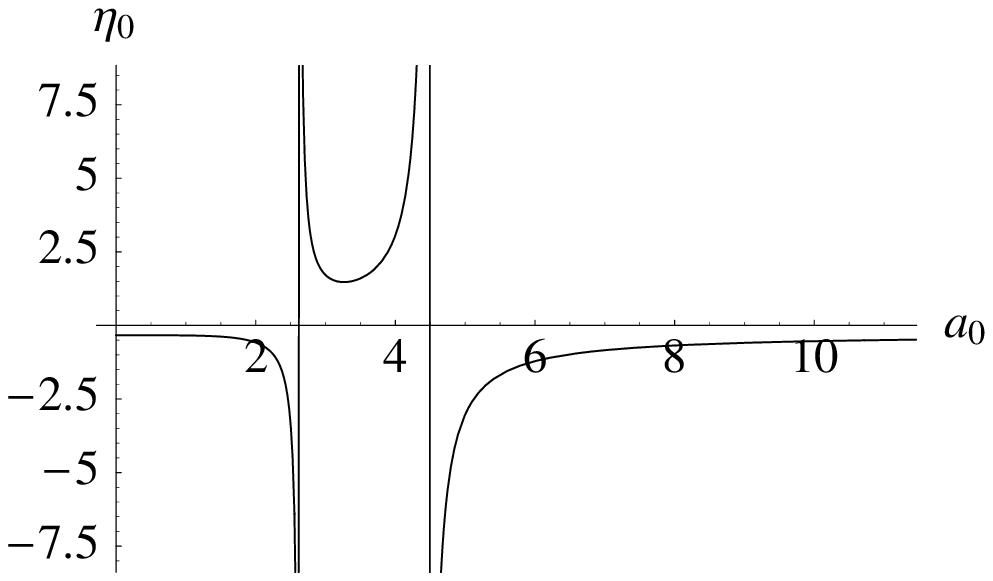} \quad
\includegraphics[width=0.25\textwidth]{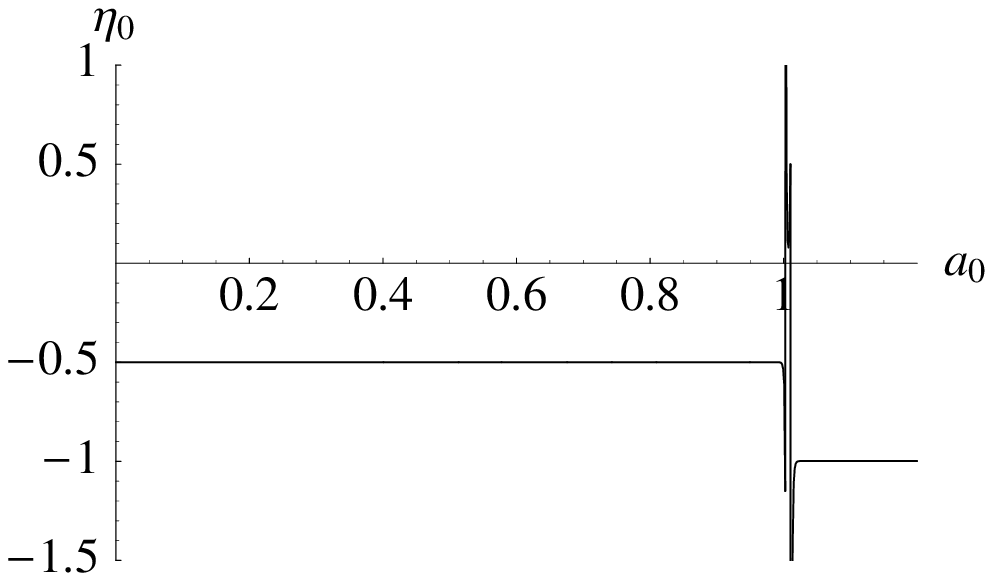}
\end{center} \centerline{(a)} \centerline{}

\begin{center}
\includegraphics[width=0.25\textwidth]{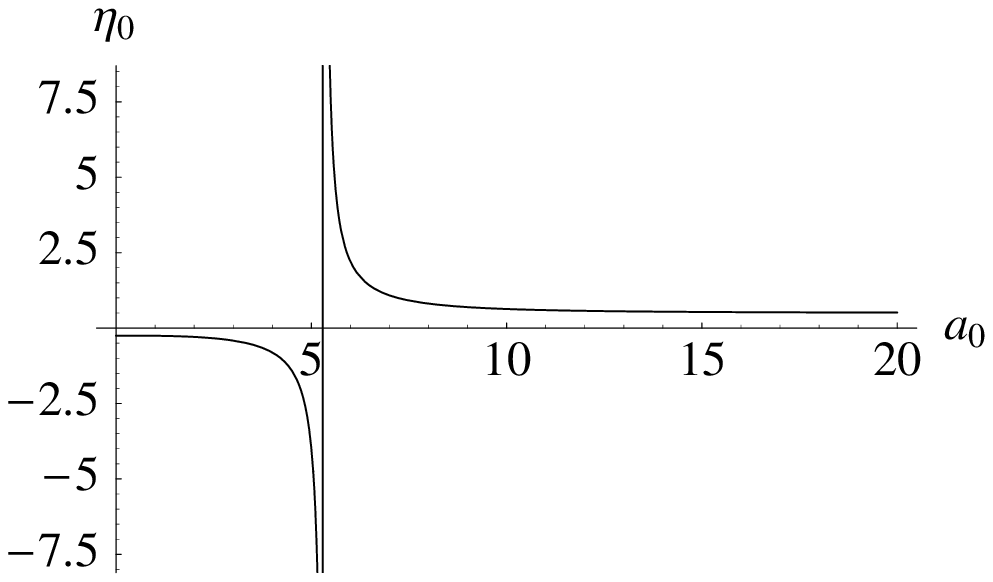} \quad
\includegraphics[width=0.25\textwidth]{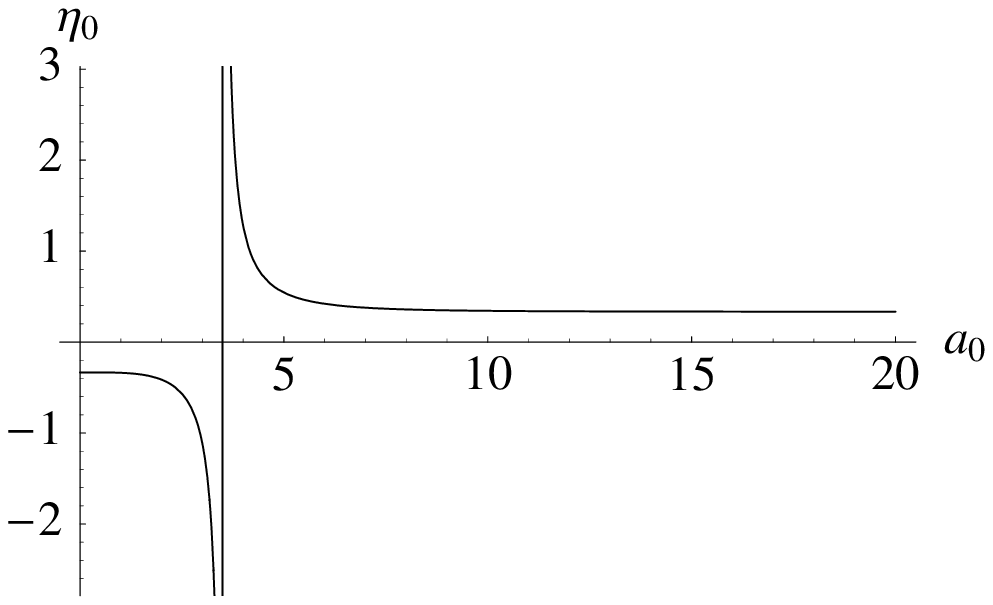} \quad
\includegraphics[width=0.25\textwidth]{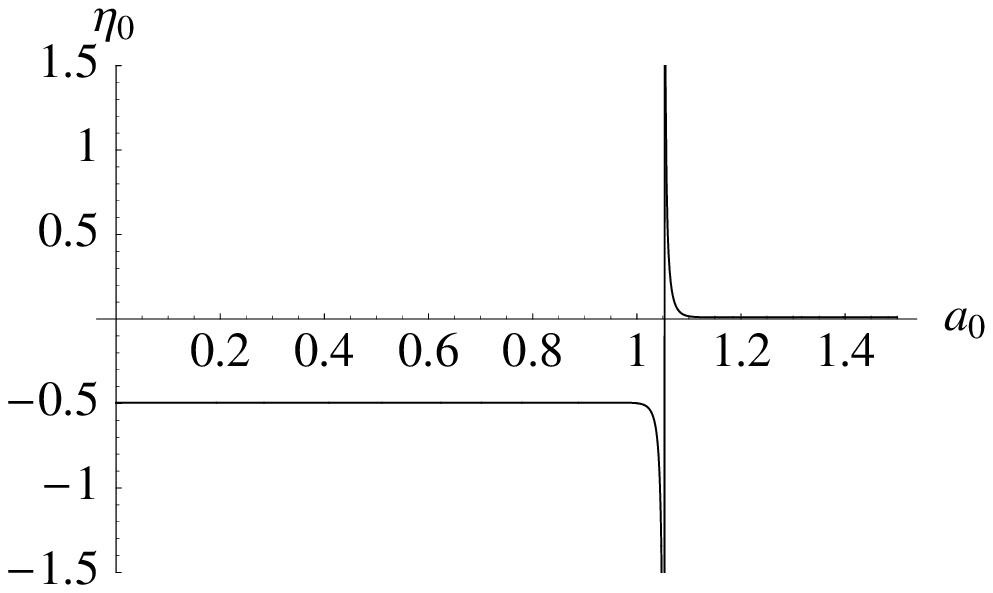}
\end{center} \centerline{(b)}

\begin{center}
\includegraphics[width=0.25\textwidth]{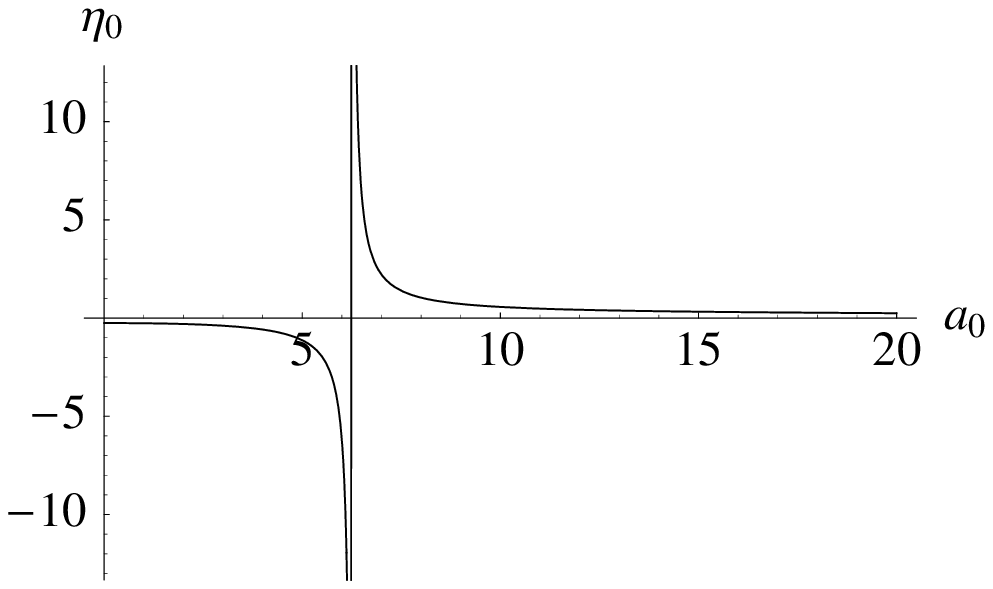} \quad
\includegraphics[width=0.25\textwidth]{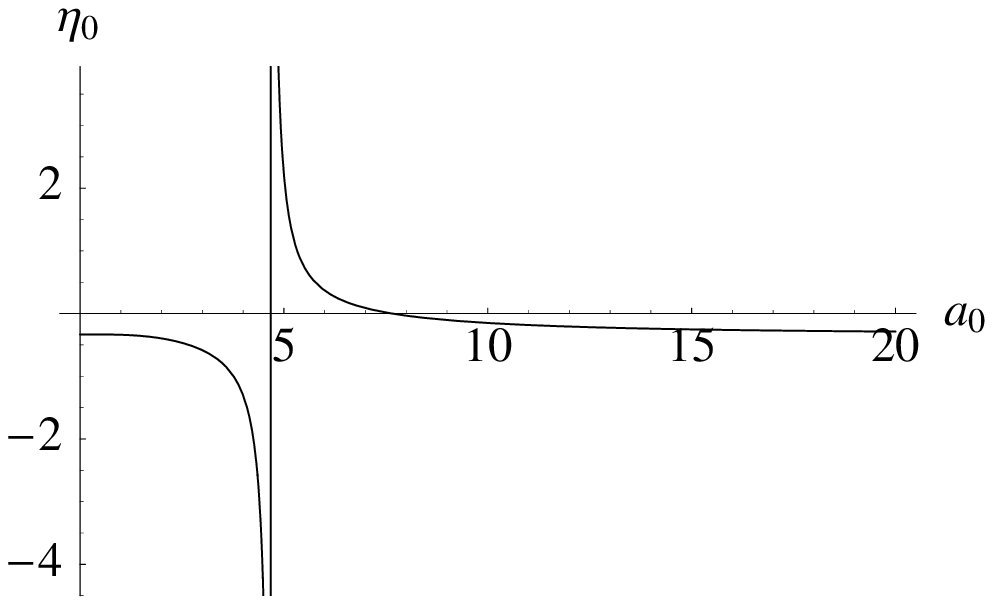} \quad
\includegraphics[width=0.25\textwidth]{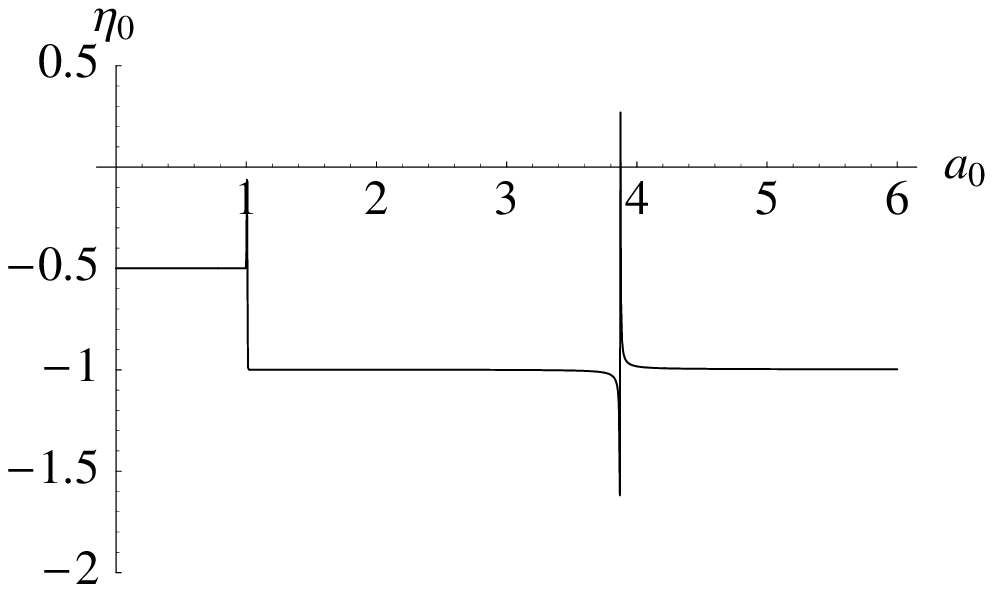}
\end{center} \centerline{(c)} \centerline{}

\caption{
These are plots for $\eta_0$ as function of $a_0$, as given on the
right hand side (rhs) of the inequalities \eq{conditions for
etab1}-\eq{conditions for etab3}.  The inequality refers to the area
above or under the curves given by the rhs of \eq{conditions for
etab1}-\eq{conditions for etab3}.  For every graphic we have $M=10$,
$Q=0$, and $\Lambda=-\frac15$, in natural units, $G=1$ and $c=1$. In
every case, from (a) to (c), the left side is for $d=4$, the center is
for $d=5$ and the right is for the limit of large $d$, illustrating
the limit $d\to\infty$. The (a) graphics are for spherical geometry,
$k=1$, the (b) are for planar geometry, $k=0$, and the (c) are for
hyperbolic geometry, $k=-1$. The physically relevant region is always
to the right of the left vertical asymptote, which marks the
gravitational radius (or the horizon)
of the vacuum
black hole solution. For (a) the region of stability is to the
right of the left asymptote, above the curve. To the right of the
right asymptote, the stability region is below the curve. For (b) and
(c), there is only the left asymptote, and the stability regions are
always above the curve. This is true for our choice of parameters,
which renders the expression \eq{conditions for etab3} the relevant
one for this particular case, because of eq{condition expression}
being positive. Note that the separation of the two asymptotes in the
spherical case, $k=1$, tends to zero as $d\to\infty$. This does not
apply to planar, $k=0,$ and hyperbolic, $k=-1$. What is notorious is
the closing of the curves to their respective asymptotes in all cases,
as the dimension increases. Also note that the higher the dimension,
the clearer the horizontal asymptotes, as $a_0$ tends to infinity. For
spherical, $k=1$, and hyperbolic geometries, $k=-1$, the horizontal
asymptote is at $\eta_0=-1$, despite the fact that for spherical
geometry the asymptote is approached from below, whereas for
hyperbolic geometry it is approached from above; for planar geometry,
$k=0$, the horizontal asymptote is at $\eta_0=0$, approached from
above.}
\label{graphics}
\end{figure}

An illustration of the above general stability conditions,
Eqs.~\eq{conditions for etab1}-\eq{conditions for etab3}
can be displayed if we put $Q=0$.
Then Eqs.~\eq{conditions for etab1}-\eq{conditions for etab3}
turn into
\begin{eqnarray}
\etab_0 <
\frac{a_0^2\left(f_0'\right)^2-2\,a_0^2\,f_0''\,f_0}{2(d-2)\,f_0
\left(-2k+\frac{(d-1)M}{a_0^{d-3}} \right)}
-\frac{d-3}{d-2}\,\,\,\,\,\, &{\rm if}& \,\,\,
 -2k+\frac{(d-1)M}{a_0^{d-3}} <0\,,
\label{conditions for etab1 Q=0}\\
\etab_0=-\infty\,, {\rm or}\,\, \etab_0=+\infty\,\,,
{\rm depending}\, {\rm on}\,
{\rm the}\, {\rm branching}\,
\,\,\,\,\,\,\,\,\, &{\rm if}& \,\,\,
-2k+\frac{(d-1)M}{a_0^{d-3}}=0\,,
\label{conditions for etab2 Q=0}\\
\etab_0 >
\frac{a_0^2\left(f_0'\right)^2-2\,a_0^2\,f_0''\,f_0}{2(d-2)\,f_0
\left(-2k+\frac{(d-1)M}{a_0^{d-3}}\right)}
-\frac{d-3}{d-2}\,\,\,\,\,\, &{\rm if}& \,\,\,
 -2k+\frac{(d-1)M}{a_0^{d-3}}>0\,.
\label{conditions for etab3 Q=0}
\end{eqnarray}
In Fig.~\ref{graphics} $\eta_0$ is plotted as a function of $a_0$ and
the regions of stability are displayed in
some chosen particular cases, namely, $d=4\,,5\,,
\infty$,$\;$ $k=1,0,-1$,$\;$ $\Lambda\neq0$, $\;$and $Q=0$.
For all plots we put further
$M=10$ and $\Lambda=-\frac15$, in natural units, i.e., $G=1$
and $c=1$. The items (a), (b), and (c) refer to the
three possible geometries of the AdS spacetime, namely, $k=1$
spherical, $k=0$ planar, $k=-1$ hyperbolic,
respectively. For each item ((a), (b), and (c)) we display three
plots, namely, for $d=4$, $d=5$, and $d=\infty$.
In the plots, the physically relevant region is always to the right of
the left vertical asymptote, which marks the gravitational radius
(or the horizon) of the
solution.  Given $d$, $k$, $M$, and $\Lambda$, Eqs.~\eq{conditions
for etab1 Q=0}-\eq{conditions for etab3 Q=0} tell us the regions in a plot
$\eta_0\times a_0$ where the stability conditions are satisfied.
Depending in which slot of the rhs of \eq{conditions for
etab1 Q=0}-\eq{conditions for etab3 Q=0} one is, the inequality on the left
hand side gives the region above or under the curves of
Fig. \ref{graphics}.

For $k=1$, Fig. \ref{graphics}(a), there are two intervals worth of
mentioning. The first interval is between the left asymptote and the
right asymptote. The region of stability is then above the curve
shown. The second interval is to the right of the right asymptote.
The stability region is given below the curve shown.  In this interval
$\eta_0<0$ for which some justification can be given,
see, e.g., \cite{poisson}.  At
the point where \eq{conditions for etab2 Q=0} holds one gets for stability
that $\eta_0=+\infty$ or $\eta_0=-\infty$ depending on the branch one
is.

For $k=0$, Fig. \ref{graphics}(b), there is only one interval worth of
mentioning. This interval is to the right of the left asymptote.  The
region of stability is then above the curve shown and $\eta_0> 0$ in
this region.

For $k=-1$, Fig. \ref{graphics}(c), there is also only one interval
worth of mentioning. This interval is to the right of the left
asymptote.  The region of stability is then above the curve shown and
$\eta_0> 0$ in this region. The parameter
$\eta_0$ can be negative in this region.

Now in each geometry (a), (b), or (c),
for finite $d$ the curves do not change qualitatively,
as can be seen displayed in the figure.
However if we take the limit $d=\infty$ we find that interesting
things happen. Firstly, the gap between the asymptotes and the curves
is reduced, and is the shorter as the dimension $d$ increases. This
shows that in this limit, the region of stability is going to be the
area limited by the respective asymptotes, both vertical and
horizontal, as the actual limiting curves given by the expressions
\eq{conditions for etab1 Q=0}-\eq{conditions for etab3 Q=0} are identified
with the asymptotes. Secondly, only for the spherical case, with our
choice of parameters, the two vertical asymptotes, limiting the two
different branches of stability, the left one given by \eq{conditions
for etab3 Q=0}, and the right one given by \eq{conditions for etab1 Q=0}, in
the limit of very large dimension, $d=\infty$, will merge,
eliminating the interval between the asymptotes, thus ending the
branch given by \eq{conditions for etab3 Q=0}. Finally, we note that in
the limit of large $a_0$, each geometry has a horizontal asymptote at
a certain value of $\eta_0$. Now, when $d$ is increased, this
horizontal asymptote approaches a certain value. In the limit
$d=\infty$, this value is $\eta_0=-1$ for spherical and hyperbolic
geometries, and $\eta_0=0$ for planar geometry (see
Fig. \ref{graphics}, right column).

\section{Conclusions}
\label{conclusions}
We have used the cut and paste procedure in order to build a class of
$d$-dimensional wormholes, with a $(d-1)$-dimensional timelike throat,
by gluing together the spacetimes of geometric-topological
$d$-dimensional charged vacuum solutions, with a negative cosmological
constant, cut somewhere above the respective gravitational
radii. After obtaining the static solutions, through the use of the
Darmois-Israel formalism, we analyzed the energy conditions, and
performed a linearized stability analysis, where the purpose was to
establish the response of the solutions to a linear perturbation
around a static configuration.  We obtained general results. Previous
results are obtainable from the present's work general results for the
appropriate choices of $d$, $k$, $\Lambda$ and $Q$.

\begin{acknowledgments}
GASD thanks Centro Multidisciplinar de {}Astrof\'{\i}sica - CENTRA for
hospitality. GASD is supported by FCT fellowship SFRH/BPD/63022/2009. 
This work was partially supported by FCT - Portugal through projects
CERN/FP/109276/2009 and PTDC/FIS/098962/2008.

\end{acknowledgments}

\end{document}